\documentclass[12pt,preprint]{aastex}

\newcommand\lsim{\lower0.5ex\hbox{$\; \buildrel < \over \sim \;$}}

\shorttitle{Complex magnetic field structures} \shortauthors{Zhang et al.}

\begin{document}

\title{Are complex magnetic field structures responsible for the confined X-class flares
in super active region 12192?}

\author{Jun Zhang\altaffilmark{}, Ting Li\altaffilmark{},
Huadong Chen\altaffilmark{}}

\altaffiltext{}{Key Laboratory of Solar Activity, National
Astronomical Observatories, Chinese Academy of Sciences, Beijing
100012, China; zjun@nao.cas.cn; hdchen@nao.cas.cn.
University of Chinese Academy of Sciences, Beijing 100049, China}

\begin{abstract}

From 2014 October 19 to 27, six X-class flares occurred in super
active region (AR) 12192. They were all confined flares and were not
followed by coronal mass ejections (CMEs). To examine the structures of
the four flares close to the solar disk center from October 22 to
26, we employ firstly composite triple-time images in each flare
process to display {the stratified structure} of these flare
loops. The loop structures of each flare in both lower (171 {\AA})
and higher (131 {\AA}) temperature channels are complex, e.g., the
flare loops rooting at flare ribbons are sheared or twisted (enwound)
together, and the complex structures have not been destroyed during
the flares. For the first flare, although the flare loop system appears as
a spindle shape, we can estimate their structure from observations,
with lengths ranging from 130 to 300 Mm, heights from 65 to 150 Mm,
widths at the middle part of the spindle from 40 to 100 Mm, and shear
angles from 16\degr~to 90\degr.
Moreover, the flare ribbons display irregular movements,
such as the left ribbon fragments of the flare on 22 swept a small
region repeatedly, and the both ribbons of the flare on 26 moved
along the same direction, instead of separating from each other.
These irregular movements also imply that the corresponding flare
loops are complex, e.g. several sets of flare loops are twisted
together. Although previous studies suggest that the background
magnetic fields prevent confined flares from erupting, we firstly
suggest based on these observations that the complex flare loop
structures may be responsible for these confined flares.

\end{abstract}


\section{Introduction}

The flares associated with coronal mass ejections (CMEs)
are termed ``eruptive flares" and
the eruptive flares usually last for a long period, from tens of
minutes to hours (e.g., Zhang et al. 2007). Some flares
are not accompanied by CMEs
in the wake of the eruption (Ji et al. 2003), these
flares are called confined flares (e.g., Wang \& Zhang 2007). The
occurrence rate of eruptive flares is dependent on the flare
intensity and duration (Kahler et al. 1989; Andrews 2003). The
fraction of flares that is associated with CMEs increases rapidly as
go from small flares to large X-class flares, reaching close to
100\% for the largest ones. For example, Yashiro et al. (2005) find
that the fraction of flares associated with CMEs increases from 20\%
for C3$-$C9 flares to 100\% for flares above X3. Wang \& Zhang
(2007) reported whether a flare is eruptive or confined is
determined by the distance between the flares and the active regions,
such as 22$-$37 Mm for eruptive flares and 6$-$17 Mm for confined
flares. On the other hand, the overlying magnetic arcades provide strong
confinement, and are
believed to play an important role in the failed eruptions (e.g.,
T{\"o}r{\"o}k \& Kliem 2005; Guo et al. 2010; Cheng et al. 2011;
Chen et al. 2013).

To explain the physical mechanism of eruptive events, many theories
and models have been proposed, in which the overlying magnetic loops
should be opened so that plasma and magnetic flux can escape
(Forbes et al. 2006). For the eruptive flares, a rising
flux rope stretches the overlying magnetic lines, then a magnetic
reconnection between the stretching lines takes place (Sturrock
1966; Masuda et al. 1994; Shibata et al. 1995; Tsuneta 1996). The
confined flares are mainly affected by the surrounding coronal
magnetic fields. T{\"o}r{\"o}k \& Kliem (2005) and Fan \& Gibson
(2007) revealed that while the overlying arcade field decreases
slowly with height, a confined event is permitted. Furthermore, the
calculations from the potential field source-surface model showed
that stronger overlying magnetic arcades will prevent energy
release, thus resulting in confined flares (Wang \& Zhang 2007; Guo
et al. 2010; Chen et al. 2015).

A large number of simulations about the question whether a
configuration fully or partly erupts have been investigated.
Examining the magnetohydrodynamic (MHD) stability, analyzing
nonlinear force-free field (NLFFF) models, and employing the
techniques of flux rope insertion (van Ballegooijen 2004; van
Ballegooijen et al. 2007) and magnetofrictional relaxation (Yang et
al. 1986), Kliem et al. (2013) confirmed that the MHD treatment of
the eruptive configuration can reappear some observed features.
Considering the condition that a toroidal flux rope embeds in a
bipolar or quadrupolar external field, catastrophe and torus
instability occur at an X-line under the flux rope where magnetic
reconnection takes place (Kliem et al. 2014a). Through studying
force-free equilibria containing two vertically arranged magnetic
flux ropes (Titov \& Demoulin 1999; Liu et al. 2012), Kliem et al.
(2014b) have demonstrated several conditions for the two ropes
activities, e.g. both the ropes turn unstable, both the ropes erupt
upward, and only the upper rope erupts while the lower rope
reconnects with the ambient flux.

In this work, we report four confined X-class flares by analyzing
the complex flare loops, with the observations from the Atmospheric
Imaging Assembly (AIA; Lemen et al. 2012) and the Helioseismic and
Magnetic Imager (HMI; Scherrer et al. 2012; Schou et al. 2012)
aboard the \emph{Solar Dynamics Observatory} (\emph{SDO}; Pesnell et
al. 2012). The data from the \emph{Interface Region Imaging
Spectrograph} (\emph{IRIS}; De Pontieu et al. 2014) are also
employed to display the special evolution of these flare ribbons.
In Section 2, we
describe the observational data. The results are presented in Section 3.
Section 4 displays the conclusions and a brief discussion.

\section{Observations}

\emph{SDO}/AIA observes the Sun in ten wavelengths
with a 0$\arcsec$.6 pixel size and a 12 s cadence. These data
reveal the solar atmospheric temperatures from $\sim$5000 K to
$\sim$20 MK. The \emph{SDO}/HMI records the full disk line-of-sight (LOS)
magnetic field with a cadence of 45 s and a spatial sampling of
0$\arcsec$.5 pixel$^{-1}$. We use the observations of 131, 171 and
1600 {\AA} to investigate the evolutions of flare loops and ribbons,
and the LOS magnetograms are
applied to display the photospheric magnetic fields of these loops.
The data adopted here were obtained from 2014 October 22 to
26 while AR 12192 was near the solar disk center. We have
de-rotated the AIA data to the same time (October 23 15:00 UT). Furthermore,
two sets of \emph{IRIS} 1330 {\AA} data are employed. The first set
of \emph{IRIS} observations was taken from 08:18 UT to 18:07 UT on
October 22, with a cadence of 33 s, a pixel scale of 0$\arcsec$.16,
and a field of view (FOV) of 120$\arcsec$${\times}$119$\arcsec$. The
second was taken from 23:01 UT on October 25 to 11:15 UT on October
26, with the same pixel scale and FOV of the first set, but the
cadence is 18 s. The 1330 {\AA} channel contains emission from the
strong C II 1334/1335 {\AA} lines that are formed in the upper
chromosphere and transition region. In order to compare the Doppler
shifts between the confined flare on October 26 and an eruptive one,
we select a set of \emph{IRIS} data for the eruptive X-class flare
on September 10, 2014. These \emph{IRIS} observations were taken
from 11:28 UT to 17:58 UT, with a FOV covering the majority of the
AR 12158 (Li \& Zhang 2015).

\section{Results}

\subsection{Overview of AR 12192}

Due to hosting the largest sunspot group since 1990, AR 12192
observed in 2014 October has been paid significant attention
(e.g., Sun et al. 2015; Thalmann et al. 2015).
According to the statistics by Chen et al. (2015), while AR 12192
passed across the visible solar disk from October18 to 29,
it produced 6 X-class and 29 M-class flares. However, only one
M-flare related to a CME (Li et al. 2015). These X-class
flares had a similar origin within the AR and common spatial and
timing characters, implying they were homologous flares
(e.g., Zhang \& Wang 2002; Sui et al. 2004; Yang et al. 2014),
and Chen et al. (2015) suggested that tether-cutting reconnection
(Moore et al. 2001) trigger these homologous flares.
Thalmann et al. (2015) provided
evidence for repeated energy release, indicating that the same
magnetic field structures were repeatedly involved in magnetic
reconnection. Sun et al. (2015) studied the magnetic conditions of
the AR and suggested that the magnetic non-potentiality over the
restriction of background field limited the eruptions. Photospheric
motions of emerged magnetic fluxes lead to shearing the associated
coronal magnetic field, which then yields a tether-cutting favorable
configuration (Chen et al. 2015).

\subsection{Complex flare loops}

One outstanding feature of AR 12192 is its poor CME production rate,
despite many X-class flares observed during its disk passage.
Previous studies suggested that the overlying background field of
the AR may play an important role in these confined flares. In this
work, we focus on the complex flare loops themselves. Four X-class
flares (from October 22, 2014 to October 26, 2014) which occurred
close to the solar disk center are studied. Figure 1 display the
loops of the first studied X-class flare (X1 in Fig. 1a) on October
22, 2014. During the flare process, the flare loops at higher
temperature (e.g. 131 {\AA}, Figs. 1d-1f, see also the online
animated version of Figure 1) were more abundant than those at lower
temperature (171 {\AA}, Figs. 1a and 1b). These loops displayed
complex structures, such as {the loops which are sheared relative to
each other} in Fig. 1e, and rooted at two flare ribbons which were
more evident at 171 {\AA} (Fig. 1a) and 1600 {\AA} (Fig. 1c). From
the corresponding LOS magnetogram (Fig. 1g), we notice that the left
ribbon overlaps the negative polarity fields, and the right ribbon
positive fields.

\begin{deluxetable}{c|llll}
\tablecaption{Spatial scales and shear angles of the flare loops (loop systems)
in X1--X4} \label{tab:mathmode}
\tablecolumns{0}
\tablenum{1}
\tablewidth{0pt}
\tablehead{
\colhead{Parameter} & \colhead{X1} & \colhead{X2} &\colhead{X3} &\colhead{X4}
}
\startdata
$Length$ (Mm) & 130 -- 300 & 173 -- 480 & 214 -- 420 & 150 -- 400 \\
$Height$ (Mm) & 65 -- 150 & 86 -- 240 & 107 -- 210 & 75 -- 200 \\
$Width$ (Mm) & 40 -- 100 &  &  &  \\
$Shear \ angles$ (\degr) & 16 -- 90 & 30 -- 85 &  &  \\
\enddata
\end{deluxetable}

To better show these flare loops, we employ
for the first time composite triple-time images of 131
{\AA} observations which can clearly display the complex loops during
flare processes. Composite triple-filter images
(e.g. Zhang et al. 2015) from SDO/AIA
are widely used to display the different temperature structures in
the solar atmosphere.
The composite triple-time images are
derived from the hint of composite
triple-filter images, and
are produced by imitating the composite
triple-filter images. The imitation is based on the observational fact that
the flare loops brighten successively from lower to higher atmospheric
levels in the flare process. In other words, each set of flare loops seems
immobile during the flare process. While we display these successively
brightened loops with different colours in ONE image,
{stratified structure} of these flare loops appears.
For example, in Fig. 1h, the blue loops are lower and shorter,
yellow loops are located at the middle layer, and the red loops are higher and longer,
and are located at the highest layer. The majority of the flare loops rooted at a
small region on the left, but
on the right the loops rooted at a long extending ribbon, thus appearing
a spindle structure (see Figs. 1c and 1h).
The lengths of flare loops can be reliably measured.
Assuming that the loops are semicircular
configurations, so the corresponding height (equals the half of the length)
can be determined also. To describe the properties of these loops, the
lengths, heights (widths) of particular loops (loop systems) of the first flare, the lengths
and heights of the other three flare loops are listed in Table 1.
{Certainly the assumption may not be well justified in complex active
regions, as the loops have a more complicated geometry far away from circular.
The heights of loops in each flare are for references only.}
To quantitatively describe the complexity of the flare loops,
shear angles in the first two flares are measured by
comparing the observed loops with the AR's prime neutral lines deduced from the magnetograms. The
flare loops are higher non-potential, e.g. strong shear (or deviating from
potential fields) in 131 {\AA}
images. The shear angles of X1 loops distribute from 16\degr~to 90\degr~(see Table 1).

Figure 3 shows the loops of the second studied X-class flare (X2 in
Fig. 3a) on October 24, 2014. Similar to the first flare displayed
in Figure 1, this flare shown also complex loop structures. The
twisted loops were observed in lower (171 {\AA},
Figs. 3b and 3c) temperature
wavelength, and the shear angles are estimated from
30\degr~to 85\degr~(see Fig. 3h and Table 1).
The last two X-class flares, which occurred on October
25 and 26 respectively, were displayed in Figure 4. Also these flare
loops were non-potential, and the shear of the loops was evident
(see Fig. 4b and the window region of Fig. 4f).
As the third and fourth flares were near the solar limb,
the widths and the shear angles of flare loop systems
can not be reliably measured. During all the four X-class flare
processes, the complex structures had not been destroyed. Also the
signal for the open and shrink of these flare loops was not evident.

\subsection{Special evolution of flare ribbons}

Previous studies for these flares have realized the large initial
separation of these flare ribbons, together with an almost absent
growth in ribbon separation, suggests a confined reconnection site
high up in the corona. Two sets of \emph{IRIS} 1330 {\AA} data with
high spatial resolution allow us to investigate the detail evolution
of the ribbons of the first and the fourth flares.  Figure 2 is time
series of the first set of \emph{IRIS} 1330 {\AA} images (also see
the online animated version of Figure 2) which show the evolution of
four fragments (marked by ``R1", ``R2", ``R3", and ``R4",
respectively) of the flare ``X1" ribbons (see Fig. 1). ``R1" which
belongs to the right ribbon and appears as a hook propagates
leftward. The moving speed of ``R1" has been deduced by examining
the evolution of the ribbons observed from IRIS images. As the
spatial resolution of the IRIS data is much high (0$\arcsec$.16 per
pixel), and the moving process lasted a long time (13.5 min, see
Figs. 2a and 2e), the speed error is small. Considering the position
error is 2 pixels while we measure the speeds, the speed error is
0.6 km s$^{-1}$, so the propagating speed is 19$\pm$0.6 km s$^{-1}$
(Fig. 2e). The three fragments ``R2", ``R3", and ``R4" belong to the
left ribbon of the flare. They appear successively, and sweep
repeatedly a small region anticlockwise. Another evolution pattern
of flare ribbons revealed by the second set of \emph{IRIS} 1330
{\AA} data is both of the two ribbons move along the same direction,
instead of separating from each other. Figures 5a and 5b show the
evolution of two ribbons of the fourth flare (X4) in the
field-of-view outlined in Fig. 4f. Both the ribbons move rightward,
with a speed of 11$\pm$0.5 km s$^{-1}$ (19$\pm$0.5 km s$^{-1}$) for
the left (right) ribbon (see Fig. 5c).

\subsection{Different Doppler shifts between a confined flare and
an eruptive one}

As there is no material escape from the solar atmosphere in a
confined flare process, the Doppler shifts between a confined flare
and an eruptive one will be different. We compare the confined flare
(X4) on 2014 October 26 with the eruptive one on 2014 September 10.
Figure 5d shows the variation of the GOES soft X-ray flux (blue
curve) in the X4 flare duration. The black cross symbols show the
Doppler shifts in the position (denoted by a yellow cross symbol in
Fig. 5b) which locates at the middle region of the two flare
ribbons. The redshift velocities (10$-$30 km s$^{-1}$) are almost
stable during the flare process.
Furthermore, Figs. 4e-4h
have already showed that these flare loops are restricted to a limited
region. For the flare with mass ejection on 2014 Sep. 10, we can not plot
truly the heights of the flare loops above the photosphere where the Doppler velocity has
been measured, as the loops at the position rise almost along the line-of-sight.
Fortunately we can track the loops which cross over the position (where
the Doppler velocity has been measured) from one footpoint to another one,
so the relative height of the loop at other position can be truly measured
(see Figs. 5e$-$5g). Near 17:25 UT on 2014 Sep. 10, the projective height of
the loops reached 100 Mm, and the rising speed exceeded 200 km s$^{-1}$.
Moreover, the eruptive X-class flare
displayed different Doppler shifts, e.g. at the
later phase of the flare, the redshift velocities in the middle of
the two ribbons increased from 30 km s$^{-1}$ to about 80$-$100 km
s$^{-1}$ (Fig. 5h, see Li \& Zhang 2015).

\section{Conclusions and Discussion}

Examining the evolution of the four flares from October 22 to 26, we
note that the loop structures of each flare in both lower (171
{\AA}) and higher (131 {\AA}) temperature channels were complex, e.g.,
the flare loops rooting at flare ribbons were twisted (enwound)
together, strong shear among loops and evidently non-potential
features of the flare loops.
\emph{IRIS} observations display that the flare ribbons underwent
irregular movement, such as the left ribbon of the flare on 22 swept
repeatedly a small region, and the both ribbons of the flare on 26
moved along the same direction, instead of separating from each
other. By comparing the confined flare X4 with the eruptive one on
2014 September 10, we find that the eruptive flare displayed a
strong Doppler redshift enhancement and the
rising speed of the flare loops reached 200 km s$^{-1}$ at the late phase of the flare,
but no redshift enhancement for the confined one, implying that the
complex flare loops did not open and then shrink. Based on these
observations, we firstly suggest that the complex flare
structures tie themselves up, and there is no enough energy to
untangle these fastened structures, so these flares are observed as confined ones.

In previous studies, the dominant idea to interpret confined flares
is the overlying loops preventing the flares from being eruptive.
Indeed there are some large-scale coronal loops above confined
flares are observed (Yang et al. 2014). However,
the complex flare loops may also play a key role, but it is always
omitted. In fact, observations reveal that there are many loops
involved in a flare, and these loops will erupt if the flare is an
eruptive one. Under the condition that these eruptive loops are
individual, e.g. potential field loops, the overlying loops should
cover fully all these individual flare loops to prevent them from
erupting. So we should observe a network (or a dome) which consists by the
overlying loops enwrapping the flare loops. On the contrary, only
one or two sets of overlying loops (arcades) are detected during
confined flare processes (Chen et al. 2013, Yang et al. 2014). These
evidences suggest that the flare loops are not individual, instead,
they are enwound together, so one or two set of overlying loops can prevent
them from erupting. Our observations supporting that the loops of
each studied flare are indeed enwound together. Liu et al. (2014)
reported a confined flare with a quasi-static cusp-shaped structure
which consists of multiple nested loops, implying that the flare
loops are also enwound together.

Another evidence to display the complex flare loops is the evolution
of flare ribbons. The flare ribbons are considered as mapping the
energy release site in solar flares. The movement of the ribbons and
its relationship to magnetic fields are important for understanding the
magnetic reconnection process. The two ribbons in many flares appear to
separate from each other during the developing process of the flares
(e.g., Fletcher \& Hudson 2002; Qiu et al. 2002; Asai et al. 2004;
Veronig et al. 2006; Miklenic et al. 2007; Temmer et al. 2007; Li \& Zhang 2009).
In our studied events, the two
ribbons of each flare are evident, but the separation of the two
ribbons is absent. Furthermore, the detail observations from
\emph{IRIS} display that the ribbons undergo special evolution. For
the first studied flare, the ribbon fragments (see Fig. 2) ``R2",
``R3" and ``R4" swept successively a small region along different
directions, indicating that the corresponding flare loops are
enwound together. AIA observations confirm that the flare loops are
enwound, e.g. shear each other in Fig. 1e. For the fourth flare,
the two ribbons move along the same direction (Figs. 5a and
5b) with speeds of 10 to 20 km s$^{-1}$, instead of separation. We
can imagine that only braiding flare loops can produce the two
ribbons moving along the same direction, and the AIA observations
(shear loops in Fig. 4) support the scenario.

During the fourth flare process, the redshift velocities (10$-$30 km
s$^{-1}$) at the middle region of the two ribbon are almost stable,
suggesting that there is no violent upward (downward) movement
enhancement in the lower atmosphere. In other words, the complex
magnetic structures relevant to the flare do not undergo ascending
(descending) process, or the complex magnetic structures are stable
and do not affect by the flare. For comparison, the eruptive X-class
flare on 2014 September 10 displayed different Doppler shifts, e.g.
at the later phase of the flare, the redshift velocities at
post-flare loop position increased from 30 km s$^{-1}$ to about 80$-$100 km
s$^{-1}$ (Fig. 5h) in the transition region (Li \& Zhang 2015),
implying the violent downward-moving cool and dense chromospheric
and transition region condensations (Fisher et al. 1985) during the
gradual phase. This study is only a preliminary step to investigate
the complex magnetic structures for confinement of solar flares. To
further evaluate the effectiveness of these complex structures, a
more robust study involving more events is needed.

\acknowledgments {This work is supported by the National Natural
Science Foundations of China (11533008 and 11673034).
The data are used courtesy of NASA/\emph{SDO} and
\emph{IRIS} science teams. }

{}

\clearpage

\begin{figure}
\epsscale{0.6} \plotone{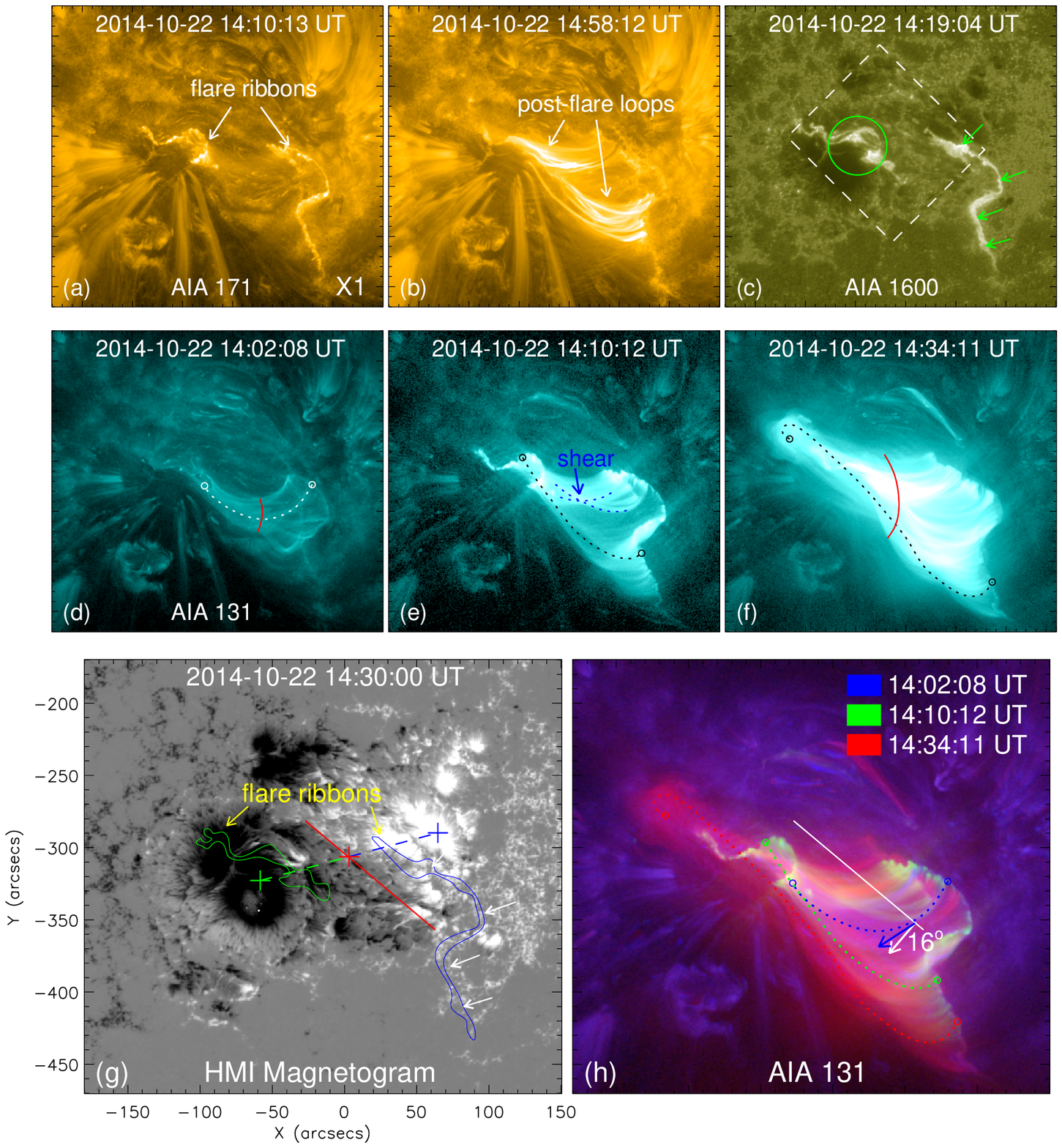}\caption{Panels (a)$-$(b): 171
{\AA} images showing the flare ribbons and post flare loops of the
first studied X-class flare (X1 in Fig. 1a) on October 22, 2014  at
lower temperature. Panel (c): 1600 {\AA} image displaying the two
flare ribbons which are denoted by one green circle and four green
arrows, respectively. The white window outlines the field-of-view
(FOV) displayed in Figure 2. Panels (d)$-$(f): 131 {\AA} images
showing the flare loops at higher temperature (also see the online
animation). White (Fig. 1d) and black (Figs. 1e and 1f) curves
outline the lengths (see Table 1) of the flare loops at different
times. Two red curves in Figs. 1d and 1f denote the widths of the
loop systems, and two crossed blue curves in Fig. 1e show the shear
of the loops. Panel (g): corresponding line-of-sight (LOS)
magnetogram. The blue and green curves are the contours of the flare
ribbons displaying in panel (c). The green, blue, and red pluses
correspond to the locations of the maximum negative flux, maximum
positive flux, and their midpoint, respectively. The red line
outlines the AR prime magnetic polarity inversion line (PIL, Chen et
al. 2015). Panel (h): composite triple-time images of 131 {\AA}
images consisting of 14:02 UT (blue), 14:10 UT (green) , and 14:34
UT (red) images. The white line is the duplication of the PIL (the
red line in Fig. 1g). The blue, yellow and red curves are the
duplications of the white curve in Fig. 1d, the black curve in Fig.
1e and the black curve in Fig. 1f, respectively. The angle
16\degr~represents the shear angle between the blue arrow
(represents the tangent of the blue curve at the intersection of the
blue curve and PIL) and the white arrow (the line perpendicular to
the PIL). An online animated version of the 131 {\AA} images is
available. The 14s animation runs from 14:00~to~15:00 UT.
\label{fig1}}
\end{figure}

\clearpage

\begin{figure}
\epsscale{1.0} \plotone{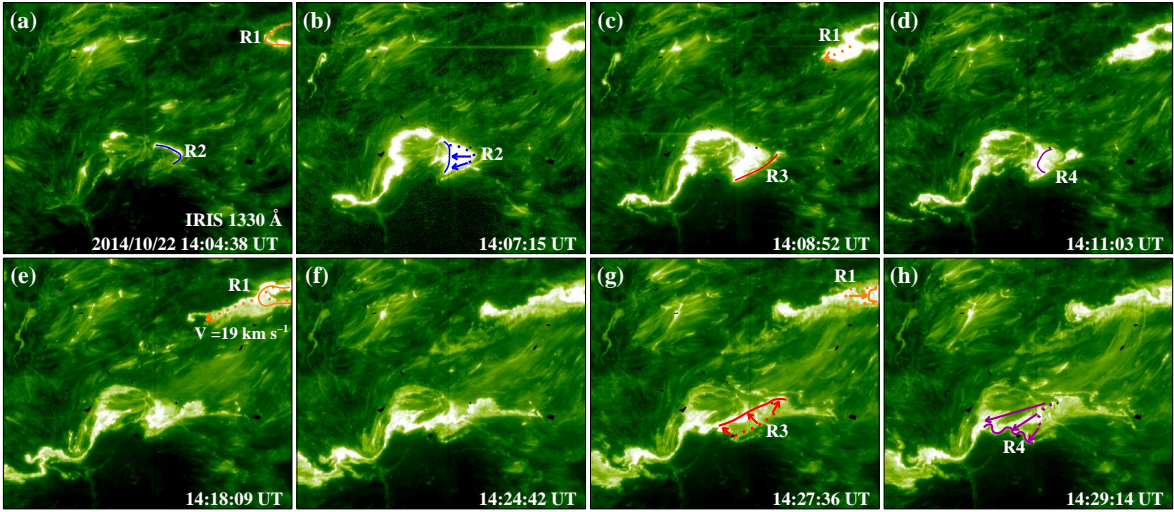}\caption{Time series of \emph{IRIS}
1330 {\AA} images showing the evolution of four fragments (marked by
``R1", ``R2", ``R3", and ``R4", respectively) of the two flare
ribbons (see Fig. 1). Panel (a): the first appearance of ``R1" (
outlines by a solid yellow curve) and ``R2" (outlines by a solid
blue curve). Panel (b): leftward sweeping of ``R2". The blue arrows
denote the sweeping direction, the blue solid curve outlines the
left boundary of ``R2", and the dotted blue curve is the duplication
of the blue curve in Fig. 2a. Panel (c): ``R3" (outlined by a red
curve) first appearance. Panel (d): ``R4" (the purple curve) first
appearance. The dotted yellow arrows in Figs. 2c and 2e denote the
propagating direction of ``R1" from the curve top of ``R1", and the
solid yellow arrow in Fig. 2g the shrinking direction of ``R1". This
propagation last almost twenty minutes (from 14:04 UT to 14:24 UT,
see Figs. 2a$-$2f), with the largest propagating speed of 19$\pm$0.5
km s$^{-1}$ (see Fig. 2e). ``R1" enlarges at first, and then shrinks
(see the solid yellow curves in Figs. 2e and 2g). The dotted curve
in Fig. 2e is the duplication of the yellow curve in Fig. 2a, and
the dotted curve in Fig. 2g is the duplication of the solid curve in
Fig. 2e. ``R3" sweeps up-leftward (see the red curve in Fig. 2g),
with the red arrows denoting the sweeping direction, and the dotted
red curve the duplicate of the curve in Fig. 2c. ``R4" sweeps
down-leftward (see the purple curve in Fig. 2h, the arrows denote
the sweeping direction). The dotted purple curve in Fig. 2h is the
duplication of the curve in Fig. 2d. An online animated version of
this Figure is available. The 9s animation runs from 14:00~to~15:00
UT. \label{fig2}}
\end{figure}

\clearpage

\begin{figure}
\epsscale{1.0} \plotone{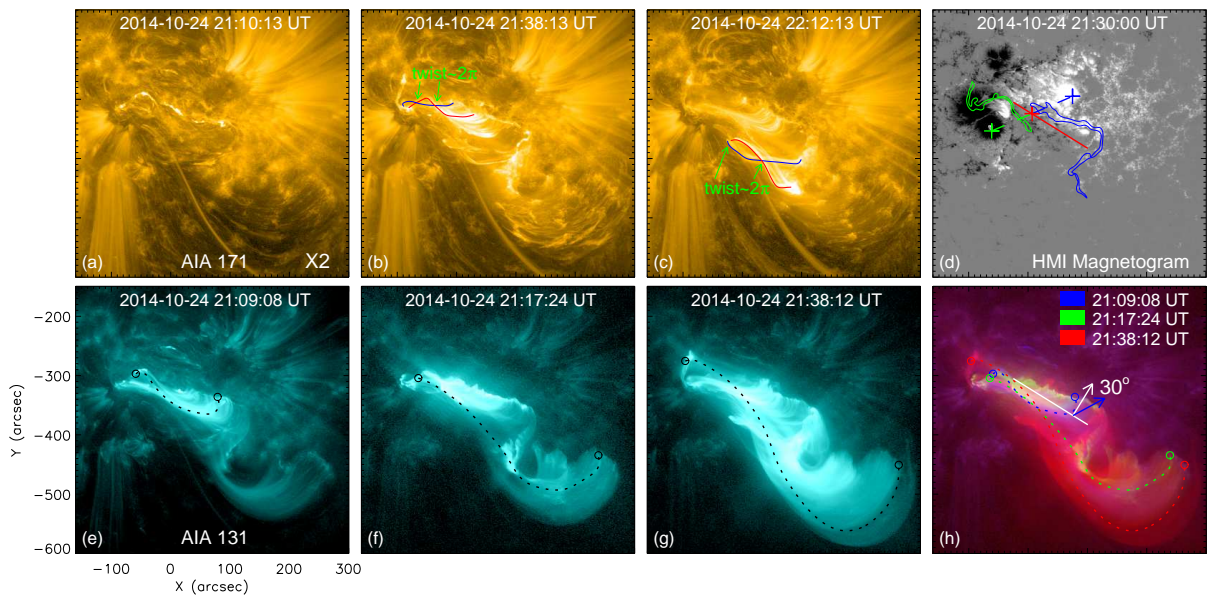}\caption{Panels (a)$-$(c): 171
{\AA} images showing the loops of the second studied X-class flare
(X2 in Fig. 3a) on October 24, 2014  at lower temperature.
The crossed red and blue curves in Figs. 3b and 3c represent the
twisted flare loops. Panel
(d): corresponding LOS magnetogram. Contours, lines
and pluses represent the same meanings as that in Fig. 1g.
Panels (e)$-$(g): 131 {\AA} images showing the loops at higher
temperature. Black curves in Figs. 3e--3g
outline the lengths (see Table 1) of the loops at different times.
Panels (h): composite triple-time images of 131 {\AA} images
consisting of 21:09 UT (blue), 21:17 UT (green) , and 21:38 UT (red)
images. Curves, lines, arrows and angle represent the same meanings
as that in Fig. 1h. \label{fig3}}
\end{figure}

\clearpage

\begin{figure}
\epsscale{1.0} \plotone{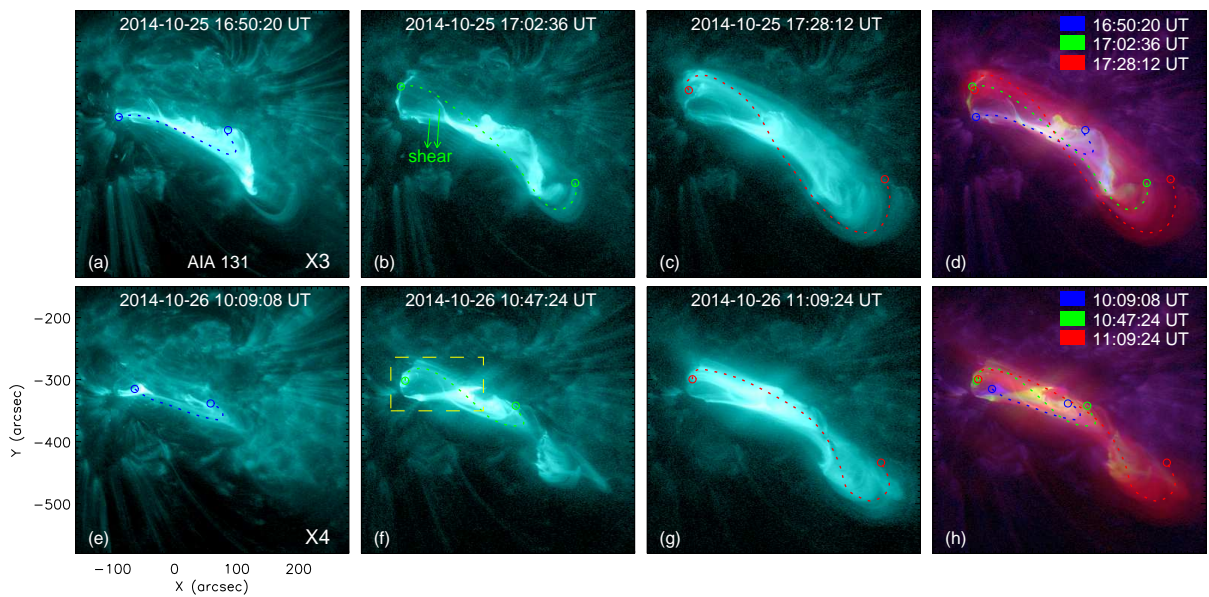}\caption{Panels (a)$-$(c): 131
{\AA} images showing the loops of the third studied X-class flare
(X3 in Fig. 4a) on October 25, 2014 at higher temperature. Blue, yellow and red
curves outline the lengths of the loops, and are duplicated to Fig. 4d.
Two yellow arrows in Fig. 4b
denote the shear loops. Panel
(d): composite triple-time images of 131 {\AA} images consisting of
16:50 UT (blue), 17:02 UT (green), and 17:28 UT (red) images. Panels
(e)$-$(g): similar to Figs. 4a--4c, 131 {\AA} images showing the
loops of the fourth studied X-class
flare (X4 in Fig. 4e) on October 26, 2014. Panel (h): similar to Fig. 4d, composite
triple-time images of 131 {\AA} images consisting of 10:09 UT
(blue), 10:47 UT (green), and 11:09 UT (red) images. The yellow
window in Fig. 4f outlines the FOV displayed in Figs. 5a and 5b.
\label{fig4}}
\end{figure}

\clearpage

\begin{figure}
\epsscale{0.6} \plotone{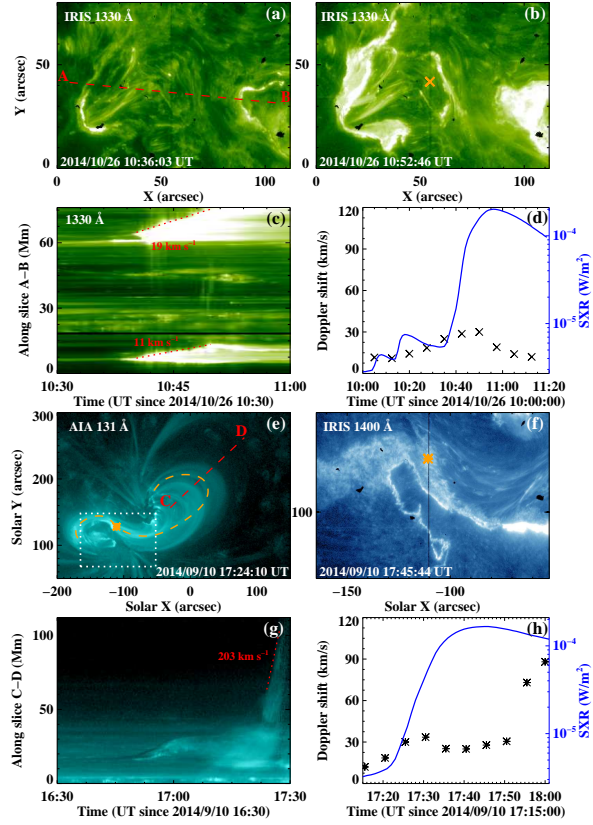}\caption{Panels (a)$-$(b): two
\emph{IRIS} 1330 {\AA} images showing the evolution of two ribbons
of the fourth flare (X4) in the FOV outlined in Fig. 4f.
Dashed line ``A-B" shows the cut position used to obtain the
stack plots shown in Fig. 5c. A yellow cross symbol in Fig. 5b denotes a
position where Doppler shifts are measured. Panel (d): the variation
of the GOES soft X-ray flux (blue curve) in the X4 flare duration.
The black cross symbols show the Doppler shifts in the position
denoted in Fig. 5b. Panel (e): AIA 131 {\AA} image showing the
loops of AR 12158 on 2014 September 10. The white window outlines
the FOV of Fig. 5f. A yellow star symbol denotes a
position where Doppler shift are
measured. Red dashed line ``C-D" shows the cut position used to obtain the
stack plot shown in Fig. 5g, and yellow dashed curve tracks a loop
system which crosses the star symbol position. The loop system erupts afterward,
with the rising speed exceeding 200 km s $^{-1}$.
Panel (f): \emph{IRIS} 1400 {\AA} image
showing a ribbon of the eruptive X-class flare in AR 12158.
A yellow star symbol denotes the same position as the star in Fig. 5e.
Panel (h): the variation of the GOES soft X-ray flux (blue
curve) in the flare (on 2014 September 10) duration. The black star
symbols show the Doppler shifts in the position denoted in
Figs. 5e and 5f. \label{fig5}}
\end{figure}

\clearpage

\end{document}